\documentclass[journal]{IEEEtran}
\usepackage{graphicx}
\usepackage{caption}
\usepackage{subcaption}
\usepackage{fullpage}
\usepackage{authblk}
\usepackage{amsmath}
\begin{document}
{\copyright\ 20xx IEEE. Personal use of this material is permitted. Permission
from IEEE must be obtained for all other uses, in any current or future
media, including reprinting/republishing this material for advertising or
promotional purposes, creating new collective works, for resale or
redistribution to servers or lists, or reuse of any copyrighted
component of this work in other works.}

\title{Tunnel Field-Effect Transistors in 2D Transition Metal Dichalcogenide Materials}

\author[1]{Hesameddin Ilatikhameneh}

\author[1]{Yaohua Tan}

\author[1]{Bozidar Novakovic}

\author[1]{Gerhard Klimeck}

\author[1]{Rajib Rahman}

\author[2]{Joerg Appenzeller}

\affil[1]{Network for Computational Nanotechnology, Department of Electrical and Computer Engineering, Purdue University, West Lafayette, IN 47907, USA}

\affil[2]{Birck Nanotechnology Center, Department of Electrical and Computer Engineering, Purdue University, West Lafayette, IN 47907, USA}
\maketitle
\section{Abstract}
In this work, the performance of Tunnel Field-Effect Transistors (TFETs) based on two-dimensional Transition Metal Dichalcogenide (TMD) materials is investigated by atomistic quantum transport simulations. One of the major challenges of TFETs is their low ON-currents. 2D material based TFETs can have tight gate control and high electric fields at the tunnel junction, and can in principle generate high ON-currents along with a sub-threshold swing smaller than 60 mV/dec. Our simulations reveal that high performance TMD TFETs, not only require good gate control, but also rely on the choice of the right channel material with optimum band gap, effective mass and source/drain doping level. Unlike previous works, a full band atomistic tight binding method is used self-consistently with 3D Poisson equation to simulate ballistic quantum transport in these devices. The effect of the choice of TMD material on the performance of the device and its transfer characteristics are discussed. Moreover, the criteria for high ON-currents are explained with a simple analytic model, showing the related fundamental factors. Finally, the subthreshold swing and energy-delay of these TFETs are compared with conventional CMOS devices. 

\section{Introduction}
Power consumption is one of the main challenges for future electronics. In this regard, tunnel FETs are among the most promising candidates for future integrated circuits (ICs) due to their small subthreshold swing (SS) and low OFF-current \cite{Appenzeller1, Appenzeller2}. Having small SS and OFF-current reduces both static and dynamic power consumption of the ICs \cite{Energy_TFET}.

One of the major drawbacks of tunnel FETs is their low ON-current. The current in TFETs is the result of band-to-band tunneling (BTBT) of the carriers. The tunneling probability is usually much smaller than 1 and as a result the ON-currents of TFETs are much smaller than those of conventional FETs. However, the tunneling probability can increase significantly if the electric field at the tunneling region is high enough. Atomically thin 2D devices are very interesting in this context for TFET applications, analogous to nanotubes \cite{Appenzeller1, Appenzeller2}, due to the tight gate control over the channel which results in high electric fields at the tunnel junction. Since the BTBT transmission probability depends on the electric field exponentially, high ON-currents are expected in 2D TFETs. However, tight gate control is not the only player in determining ON-currents and there are other critical factors, such as band gap, effective mass (m$^*$), and doping concentration. 


The well known scaling length theory \cite{Scaling1} can be used to quantify the effect of gate control on the electric field at the tunnel junction \cite{Scaling_TFET}. This theory provides a simple analytic way to understand how various device parameters affect the spatial variation of the potential along the channel \cite{Scaling1} described by a modified 1D Poisson equation,
\begin{equation}
\label{eq:scaling1}
\frac{d^2V}{dx^2}-\frac{V}{\lambda^2} =0 \\
\end{equation}
where $V$ and $\lambda$ are the electrostatic potential and the natural scaling/decay length of the potential, respectively. In double gated FETs, $\lambda$ is given by \cite{Scaling2}
\begin{equation}
\label{eq:scaling2}
\lambda=\sqrt{  \frac{\epsilon_{ch}}{2\epsilon_{ox}} (1+ \frac{\epsilon_{ox}}{4\epsilon_{ch}} \frac{t_{ch}}{t_{ox}}) t_{ch} t_{ox} } \\
\end{equation}
where $\epsilon_{ch}$, and $\epsilon_{ox}$ are the dielectric constants of channel and oxide respectively, while $t_{ch}$, and $t_{ox}$ are their thicknesses. According to equation (\ref{eq:scaling2}), reducing the channel thickness reduces the natural scaling length of the potential which results in higher electric fields and ON-currents. 

Besides advantageous 2D electrostatics, there are other incentives for using 2D materials for future electronics \cite{jena2D}. For example, thinning the 3D material based FETs (which is required in transistor scaling) {increases the effect of surface roughness on carrier transport which leads to lower mobility} \cite{roughness}. However, 2D materials have {relatively} weak inter-layer bonds and can be exfoliated easily without surface roughness \cite{VdW1, VdW2, Eg, MoS2_Nature}. Moreover, there are no dangling bonds in 2D TMDs unlike thinned 3D materials \cite{dangling1, dangling2, dangling3}. Another advantage of 2D materials is that thinning the material does not increase the band gap of the material as much as it does in 3D materials \cite{Eg_d_MoS2, Eg}. This is again due to the weaker coupling between stacked layers in 2D materials. For example, thinning InAs nanowires to achieve better gate control can increase the band gap more than 100\% \cite{Eg_d_InAs}, while the increase in band gap from bulk to monolayer is much smaller in 2D materials \cite{Eg_d_MoS2}. This is useful in the TFETs as a larger badgap results in a lower ON-current. Moreover, mono-layer 2D materials usually have small dielectric constants \cite{eps_tmds}, which can also increase the ON-current and reduce the drain induced barrier lowering (DIBL) in TFETs. 

Previously, TMD TFETs have been simulated without solving the Poisson and the transport equations self-consistently \cite{nonscf1, nonscf2}. In these works, the electric field at the source-channel junction has been approximated assuming that the 1D Poisson equation (equation (\ref{eq:scaling1})) is accurate throughout the entire source-to-channel junction. Although $\lambda$ is a critical factor in determining the electric field, it is not the only factor. Other factors such as the depletion width in the source region are important too. The non self-consistent results predict very large ON-currents for all TMD TFETs irrespective of the channel material. In this paper, the correctness of these assumptions for TMD TFETs has been investigated by solving the 3D Poisson equation self-consistently with full band quantum transport, and the factors limiting the ON-currents are elucidated. 

The band structure and electronic properties such as band gap, m$^*$, and dielectric constant of TMD materials depend on the number of layers. Consequently, devices with different number of layers show different characteristics. Although increasing the thickness reduces the band gap, it increases $\lambda$ and decreases the electric field at the source-to-channel junction. Moreover, multi-layer TMDs usually exhibit indirect band gaps which implies that phonons need to be involved in the BTBT process. As a result, it is favorable to use mono-layer TMDs and instead explore different materials to obtain small values of $\lambda$, direct band gap, and m$^*$ simultaneously. Since TMDs form a general class of materials of the form MX$_2$, where M is a transition metal (Mo, W, etc.) and X is a Chalcogenide (Te, Se, S), a variety of material parameters can be accessed by the correct choice of material. The field of 2D materials is still at its infancy as novel materials are being discovered \cite{black_phosphorus, silicene}, which opens up opportunities for TFET designs. Accordingly, in this paper, only mono-layers of a set of more common TMDs, whose critical parameters span the design space of TFETs, are studied.

\section{Simulation method}

\begin{figure}[h!]
        \centering
        \begin{subfigure}[b]{0.35\textwidth}
                \includegraphics[width=\textwidth]{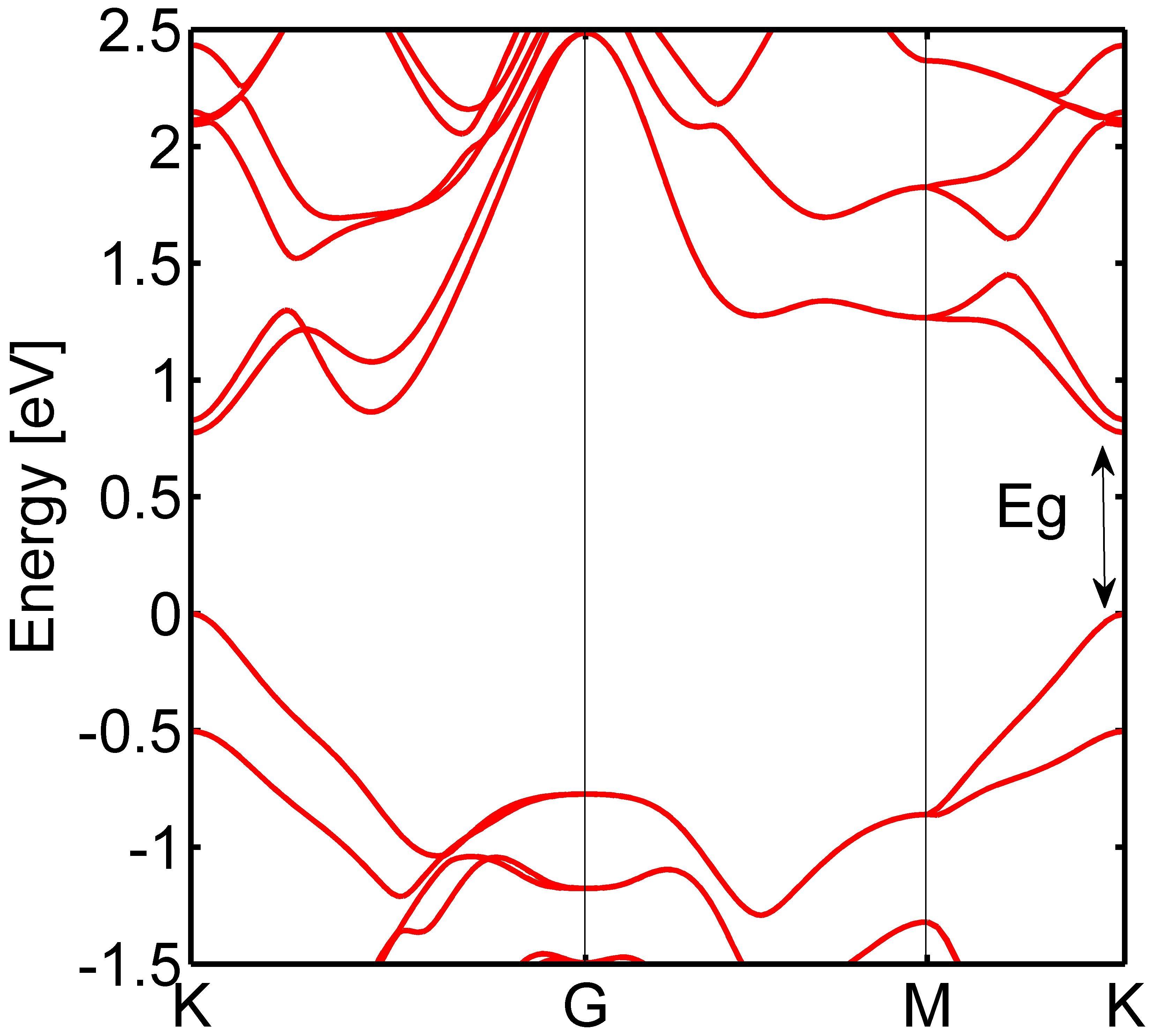}
        \end{subfigure}%
        \caption{Bandstructure of a mono-layer WTe$_2$.}\label{fig:EK_WTe2}
\end{figure}

The TMD Hamiltonian is represented by an sp$^3$d$^5$ 2nd nearest neighbor tight-binding (TB) model with spin-orbit interaction. The Slater-Koster TB \cite{slater_koster} parameters are optimized based on first principles bandstructures obtained from density functional theory (DFT) with the generalized gradient approximation (GGA). DFT-GGA has been shown to provide band gaps and effective masses in TMDs comparable to experimental measurements \cite{dft_tmdc}. The motivation for using a DFT guided TB model is that a realistically extended device size can be simulated at ease compared to computationally expensive and size limited ab-initio methods \cite{DFTMapping_YaohuaTan}. Bandgaps and effective masses of mono-layer MoS$_2$, WSe$_2$, MoTe$_2$, and WTe$_2$ obtained from our TB model are listed in table \ref{tab:tmd_prop}. The TB parameters are general and capture the bandstructure of both bulk and monolayer TMDs. {As an example, the TB bandstructure of monolayer WTe$_2$ is shown in Fig. \ref{fig:EK_WTe2}.}

In this work, self-consistent Poisson-QTBM (Quantum Transmitting Boundary Method \cite{qtbm1}) methodology has been used within the tight binding description. The QTBM method is equivalent to the well known non-equilibrium Green's function (NEGF) approach without scattering, but is a more computationally efficient implementation \cite{qtbm2}. In this method, the Schroedinger equation with open boundaries {is solved using the following equation}
\begin{equation}
\label{eq:qtbm1}
(EI - H -\Sigma) \psi_{S/D}= S_{S/D} \\
\end{equation}
{where $E$, $I$, $H$, and $\Sigma$ are energy, identity matrix, device Hamiltonian, and total self energy due to open boundaries and $\psi$ and $S$ are the wave function and a career injection term from either source (subscript S) or drain (subscript D) \cite{qtbm2}. The electron and hole carrier density and current can be obtained from the wave function $\psi$ \cite{qtbm2}. Since in-plane and out-of-plane dielectric constants ($\epsilon^{in}$ and $\epsilon^{out}$) of TMD materials are different, the Poisson equation reads as follows if considering the z direction to be along the out-of-plane direction (or c-axis) of the TMDs }
\begin{equation}
\label{eq:poisson3d}
\frac{d}{dx} (\epsilon^{in}  \frac{dV}{dx} )+\frac{d}{dy} (\epsilon^{in}  \frac{dV}{dy})+\frac{d}{dz} (\epsilon^{out}  \frac{dV}{dz})=-\rho\\
\end{equation}
{where $V$ and $\rho$ are the electrostatic potential and total charge, respectively. The dielectric constant values ($\epsilon^{in}$ and $\epsilon^{out}$) for TMD materials are taken from ab-initio studies \cite{eps_tmds}, \cite{eps_MoS2} and are listed in table \ref{tab:tmd_prop}.} In this work, transport simulations have been performed with the nanodevice simulation tool NEMO5 \cite{nemo5_1, nemo5_2}.

\begin{table}[h!]\center
\caption{\label{tab:tmd_prop} {{Band gap (Eg), electron and hole effective masses ($m^*_e$ and $m^*_h$) and in-plane and out-of-plane relative dielectric constants ($\epsilon_r^{in}$ and $\epsilon_r^{out}$) of TMD materials. All the parameters are obtained from TB bandstructure of TMD monolayers with the exception of the dielectric constants which are taken from ab-initio studies \cite{eps_tmds}, \cite{eps_MoS2}.}}}
    \begin{tabular}{| l | l | l | l | l |}
    \hline
	Parameters & MoS$_2$ & WSe$_2$ & MoTe$_2$ & WTe$_2$ \\ \hline
	Eg [eV] &	1.68 & 1.56 & 1.085 & 0.75 \\
	$m^*_e$ [m$_0$] & 0.52 & 0.36 & 0.57 & 0.37 \\
	$m^*_h$ [m$_0$] & 0.64 & 0.5 & 0.75 & 0.3 \\
	$\epsilon_r^{in}$ & 4.2 & 4.5 & 8 & 5.7 \\
	$\epsilon_r^{out}$ & 2.8 & 2.9 & 4.4 & 3.3 \\
	\hline
    \end{tabular}
\end{table}

\section{Results and discussion}
\begin{figure}[h!]
        \centering
        \begin{subfigure}[b]{0.45\textwidth}
                \includegraphics[width=\textwidth]{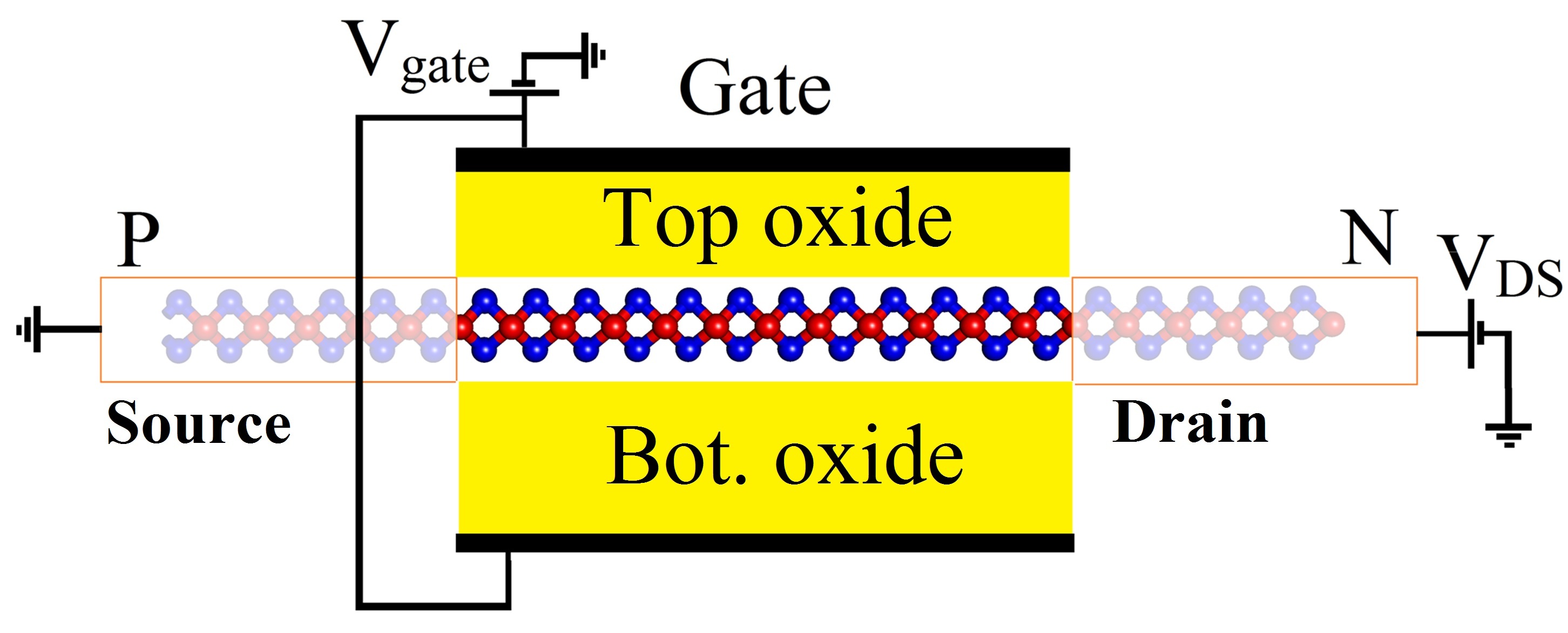}
                \label{fig:struct1}
        \end{subfigure}%
        \caption{Physical structure of a mono-layer TMD TFET with channel and source/drain lengths of 15nm and 10nm, respectively. Source and drain regions have a doping level of 1e20 cm$^{-3}$ and EOT is fixed to 0.43 nm.}\label{fig:struct}
\end{figure}

All simulated TMD TFET devices assume a structure as shown in Fig. \ref{fig:struct}, and have channel and source/drain lengths of 15nm and 10nm, respectively. A doping level of 1e20 cm$^{-3}$ is assumed in the source and drain regions which {seems feasible by molecular doping of source and drain contact regions \cite{Q1e14}}. A source-drain voltage V$_{DS}$ of 0.5V is used unless mentioned otherwise. Equivalent oxide thickness (EOT) of top and bottom oxides is set to 0.43 nm to be consistent with International Technology Roadmap for Semiconductors (ITRS) projections for 2027 \cite{ITRS2}. 

\subsection{Transfer characteristics of TMD TFETs}
\begin{figure}[h!]
        \centering
        \begin{subfigure}[b]{0.45\textwidth}
                \includegraphics[width=\textwidth]{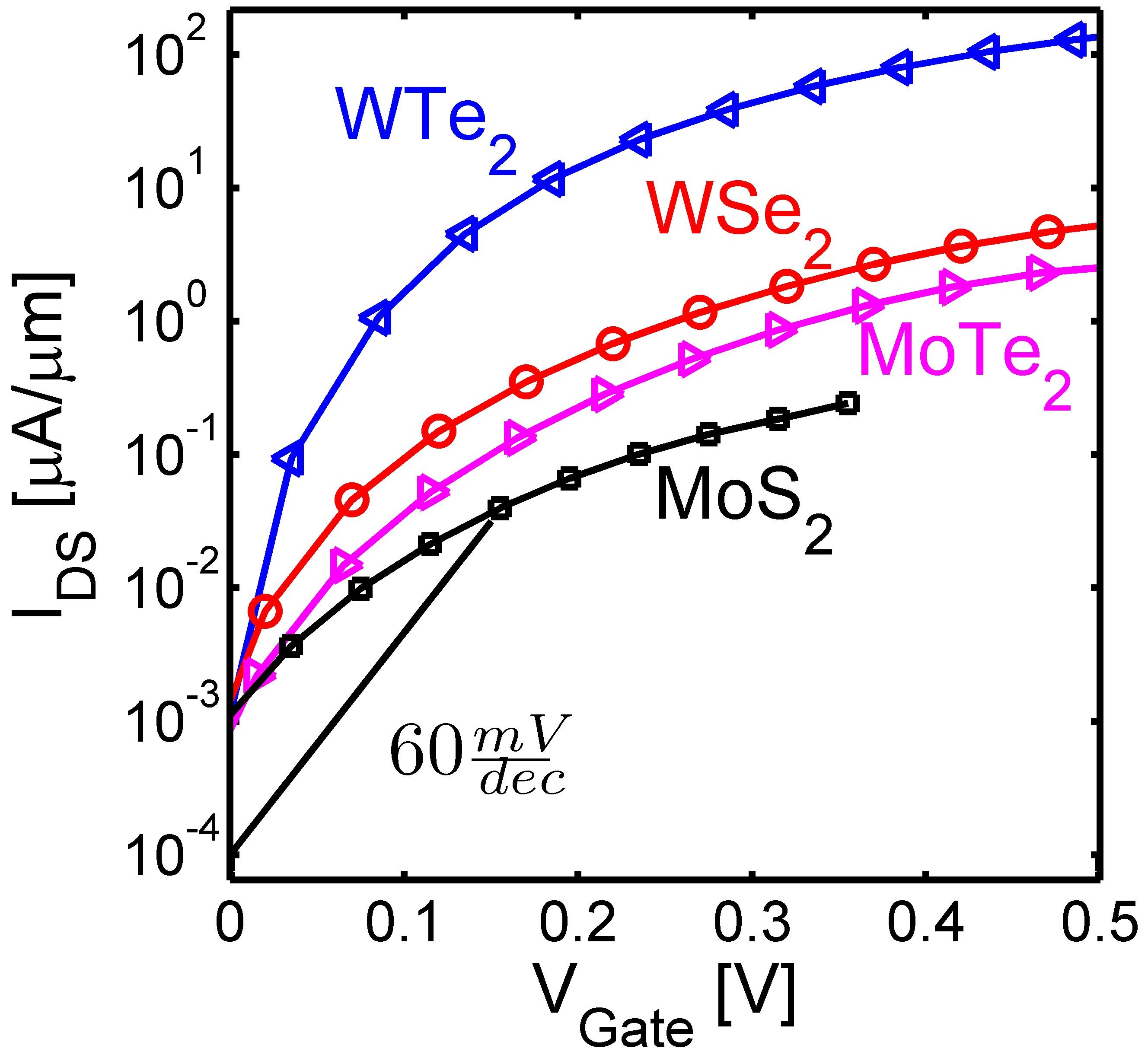}
                \label{fig:IdVg1}
        \end{subfigure}%
        \caption{Transfer characteristics of TMD TFETs with $I_{OFF}$ = 1$nA/\mu m$.}\label{fig:IdVg}
\end{figure}

Fig. \ref{fig:IdVg} shows the transfer characteristics of the TMD TFETs with OFF-current fixed to 1 $nA/\mu m$ at 0 gate voltage. It is worthwhile to notice that in TFETs lower OFF-currents can be readily achieved without losing too much ON-current. This is because the slope of the $I_{DS}-V_{Gate}$ curve is very steep in the low current regime. As a result, a much lower OFF-current can be obtained with a very small reduction of $V_{Gate}$.
For example, an OFF-current of 0.1 $nA/\mu m$ (10 times smaller than before) can be obtained with the ON-current just being approximately 5-10\% smaller.

The simulation results show that WTe$_2$ TFETs can provide highest performance in terms of ON-current and SS in comparison to the other TFETs. Since WTe$_2$ has the smallest band gap and effective mass compared to the other TMDs, its ON-current is significantly higher. Notice that despite the fact that MoTe$_2$ has a smaller band gap than WSe$_2$, it shows a smaller current. The values of ON-current (the current at $V_{GS}$=$V_{DS}$=$V_{DD}$=0.5V), band gap and reduced effective masses ($1/m^*_r = 1/m^*_e + 1/m^*_h$) of these TMDs are listed in table \ref{tab:tfet_param}. {Although MoS$_2$ has a high effective mass for tunneling applications, it is ideal for ultra-scaled MOSFET applications where a high effective mass can suppress the source to drain tunneling \cite{sub12, lake}.}
\begin{table}[h!]\center
\caption{\label{tab:tfet_param} {ON-current $\rm I_{ON}$ [$\mu A/ \mu m$], band gap Eg [eV], reduced effective masses $m^*_r$ [m$_0$], natural scaling length $\lambda$ [nm], \textit{effective} natural scaling length $\Lambda$ [nm] from simulation and $\eta$ values [unitless] of TMD TFETs}}
    \begin{tabular}{| l | l | l | l | l | l | l |}
    \hline
	Material & $I_{ON}$  & Eg & $m^*_r$ & $\lambda$ & $\Lambda$ & $\eta$ \\ \hline	
	WTe$_2$ & 127 & 0.75 & 0.17 & 0.45 & 2.45 & 3.15\\
	WSe$_2$ & 4.6 & 1.56 & 0.21 & 0.41 & 2.5 & 5.2\\
	MoTe$_2$ & 2.3 & 1.08 & 0.32 & 0.5 & 2.7 & 5.8\\
	MoS$_2$ & 0.3 & 1.68 & 0.29 & 0.38 & 2.5 & 6.3 \\	
	\hline
    \end{tabular}
\end{table}
To understand the origin of the difference between ON-currents of these devices, one needs to consider several factors: electron and hole effective masses, band gap, body thickness, EOT of oxide, source-to-channel potential difference, source and drain doping level. These factors are considered in the analytic equation for the current in the ON-state of a TFET \cite{Seabaugh1}:
\begin{equation}
\label{eq:Ion1}
I \propto \textrm{exp}\left(\frac{-4 \sqrt{2m_r^*} E_g^{3/2}} {3q \hbar E} \right)	 \\
\end{equation}
where $q$, $\hbar$, and $m_r^*$ are the charge of an electron, the reduced Plank constant, and the reduced effective mass, respectively. E is the electric field at the source-channel junction and can be approximated by $(\phi_S-\phi_{ch})/\Lambda$, where $\Lambda$ is the \textit{effective} natural scaling length of the potential at the source-channel junction and $q(\phi_S-\phi_{ch})$ is the source-channel potential difference. At threshold voltage, $q(\phi_S-\phi_{ch})$ equals $E_g$. For small overdrive voltages $(V_{GS}-V_{Th})$, the current can be simplified using $q(\phi_S-\phi_{ch}) \approx E_g$  to
\begin{equation}
\label{eq:Ion2}
I \propto \textrm{exp}\left(\frac{-4 \Lambda \sqrt{2m_r^* E_g} } {3 \hbar} \right)	 \\
\end{equation}

This equation shows the fundamental factor in determining the current in TFETs is
\begin{equation}
\label{eq:Ion3}
\eta = \frac{\Lambda \sqrt{m_r^*  E_g}}{\hbar} \\
\end{equation}
The factor $\eta$ depends on the device design (EOT, body thickness, and doping) through $\Lambda$ and band structure of channel through $m_r^*$ and $E_g$. Note that for small overdrive voltages, the band gap and the effective mass have an equal impact on the current. For example, despite the fact that MoTe$_2$ has a smaller band gap compared with WSe$_2$, it has a larger reduced effective mass which can partly compensate for the reduction in the band gap. 

The parameter $\Lambda$ is composed of two components: the depletion width in the source region ($W_D$) and the natural scaling length ($\lambda$) in the gated region. The difference between this analysis and previous works \cite{nonscf1, nonscf2, Das} is that the \textit{effective} natural scaling length $\Lambda$ is calculated from a self-consistent potential profile which includes the effect of both $W_D$ and $\lambda$ accurately. Ideally, the dielectric constant of the channel should be small to make $\Lambda$ as small as possible. $W_D$ and $\lambda$ depend on the in-plane and out-of-plane dielectric constants, respectively. {The difference between the values of $\Lambda$ and $\lambda$, listed in table \ref{tab:tfet_param}, shows the importance of $W_D$.} According to DFT simulations, MoTe$_2$ has the higher dielectric constant if compared to WSe$_2$ \cite{eps_tmds} which further lowers its ON-current.
The factor $\eta$ for these materials is shown in table \ref{tab:tfet_param}. The larger the $\eta$, the smaller the current. By comparing the $\eta$ values, it can be seen that WSe$_2$ provides higher currents close to the threshold voltage compared to MoTe$_2$ TFETs in spite of its higher bandgap. Moreover, it is expected that WTe$_2$ TFET produces the highest ON-current. Note that by employing equation (\ref{eq:Ion2}), the ON-current ratios between TMD FETs from different materials can be reproduced within reasonable accuracy when using the $\eta$ values from table \ref{tab:tfet_param}. 

\begin{figure}[h!]
        \centering
        \begin{subfigure}[b]{0.25\textwidth}
                \includegraphics[width=\textwidth]{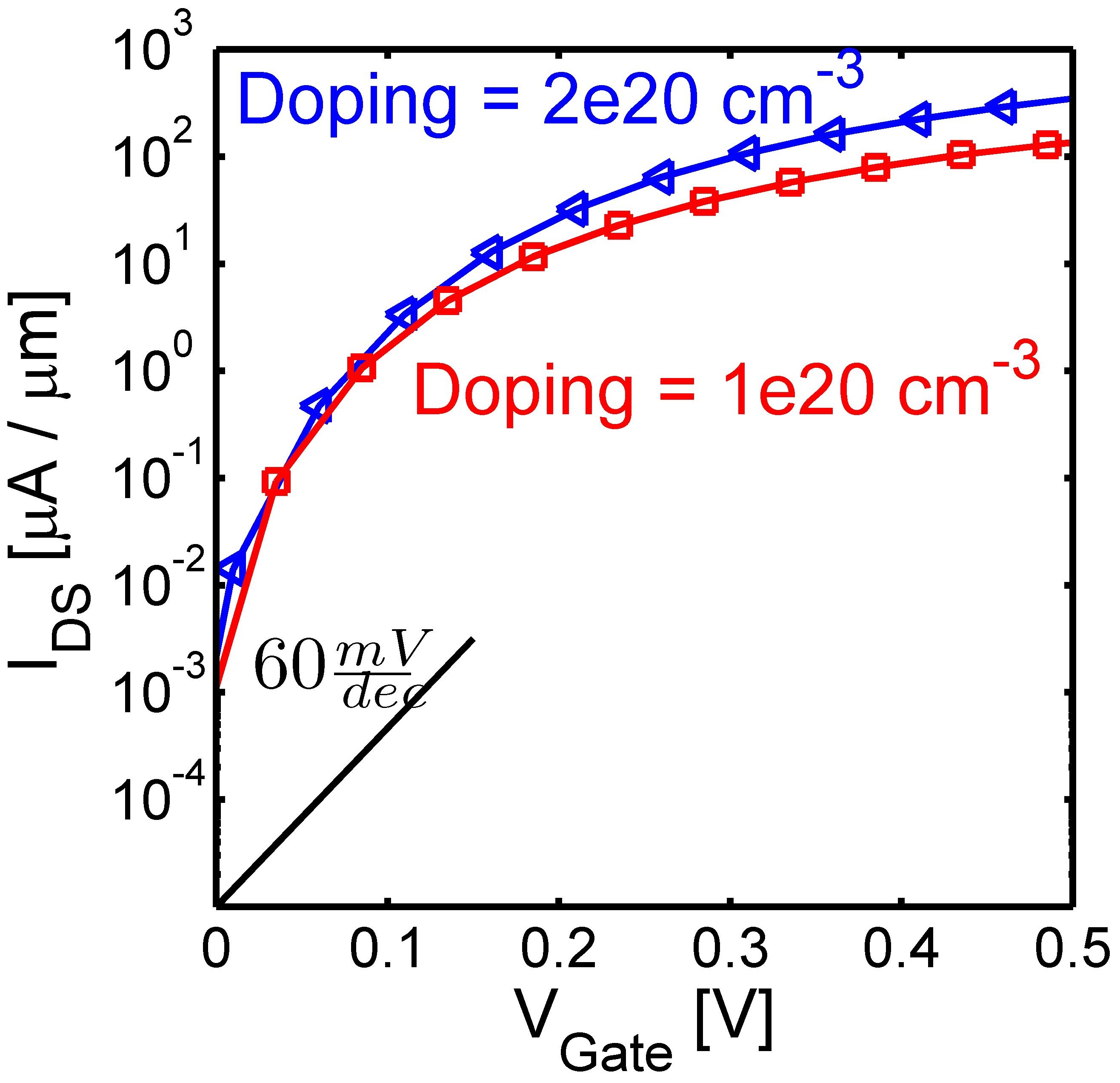}
                \caption{}
                \label{fig:IdVg_dop}
        \end{subfigure}%
        ~ 
        \begin{subfigure}[b]{0.24\textwidth}
                \includegraphics[width=\textwidth]{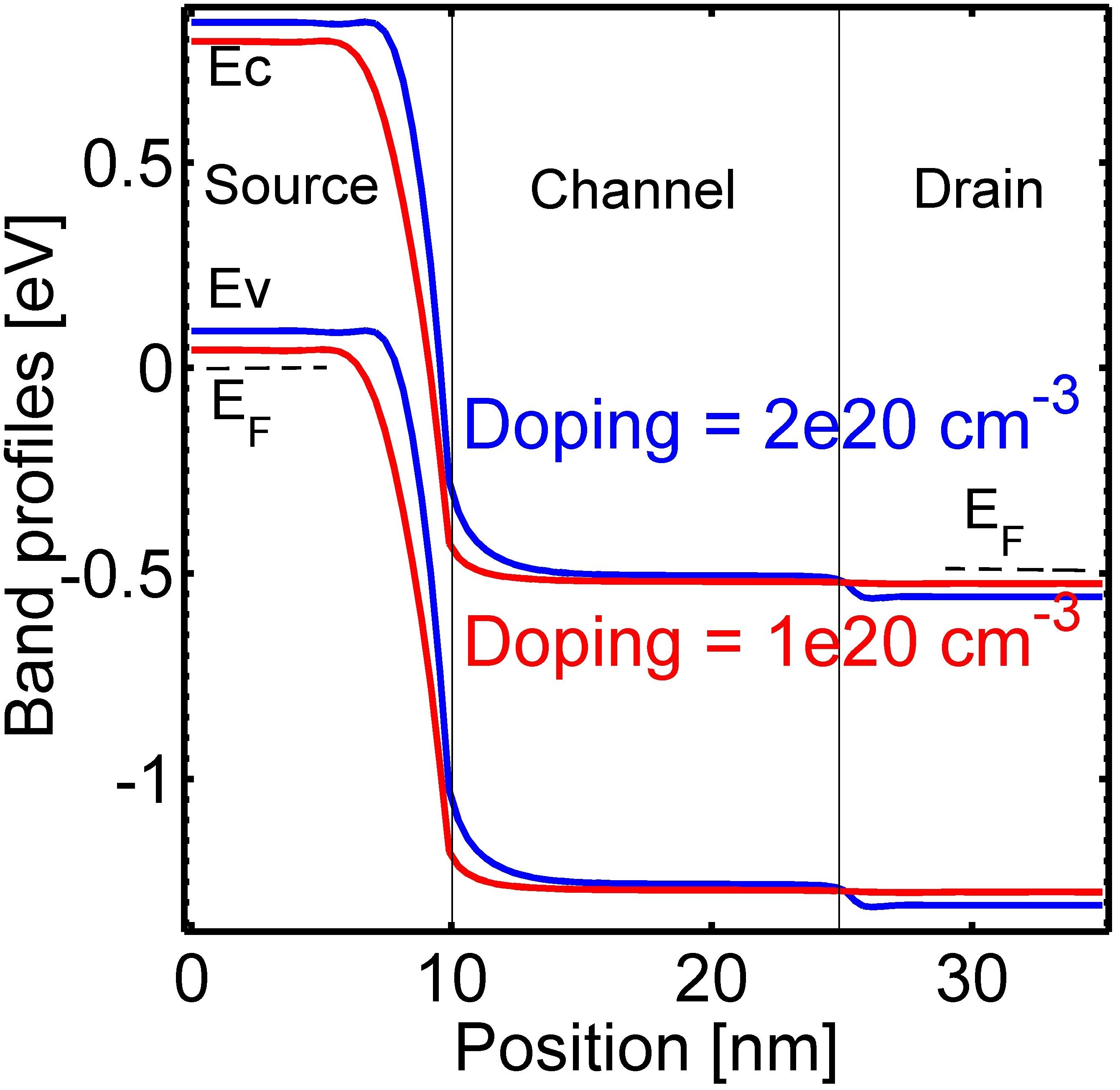}
                \caption{}
                \label{fig:Pot_dop}
        \end{subfigure}
        ~ 
        \caption{a) Transfer characteristics and b) band diagrams of WTe$_2$ with doping levels of 1e20 cm$^{-3}$ and 2e20 cm$^{-3}$. }\label{fig:dop_effect}
\end{figure}

From Fig. \ref{fig:IdVg}, one can conclude that WTe$_2$ TFETs show promising ON-currents and a steep SS compared to other TMD TFETs. It is possible to further increase the ON-current of WTe$_2$ TFET by increasing the doping concentration at the source and drain regions. Increasing the doping decreases the depletion width of the source-to-channel interface, which also decreases $\eta$ according to  equation (\ref{eq:Ion3}). Consequently, the ON-current increases. 
Fig. \ref{fig:IdVg_dop} shows a comparison between the ON-currents of WTe$_2$ TFETs with source/drain doping levels of 1e20 cm$^{-3}$ and 2e20 cm$^{-3}$. In the case of 2e20 cm$^{-3}$ doping level, the ON-current increases to 350 $\mu A/\mu m$ which is a very high ON-current if compared with other TFETs. It is important to notice that the electric field at the source-to-channel interface depends on both the natural scaling length $\lambda$ and the depletion width $W_D$. As a result, reduction of $\lambda$ is not sufficient for high ON-currents. A smaller $W_D$ is also needed. In this regard, the doping of the source and drain region should be designed carefully, whether this doping is chemically or electrically induced. Fig. \ref{fig:Pot_dop} shows the band profiles for these two different doping concentrations. It is apparent that the higher doping concentration shows higher electric field due to the smaller $W_D$. {Notice that the drain region is usually doped less than the source region to suppress the p-branch in the I-V characteristics and improve the OFF-state performance of the TFET \cite{ldd}. However, due to a relatively large band gap and effective mass of TMD materials, the minimum achievable current in TMD TFETs is always very small and below 1 $nA/\mu m$ even for a drain doping level of 2e20 cm$^{-3}$. Consequently, lowering the drain doping does not have any significant impact on the transfer characteristics of TMD TFETs.}

\subsection{C-V and DIBL}

\begin{figure}[h!]
        \centering
        \begin{subfigure}[b]{0.4\textwidth}
                \includegraphics[width=\textwidth]{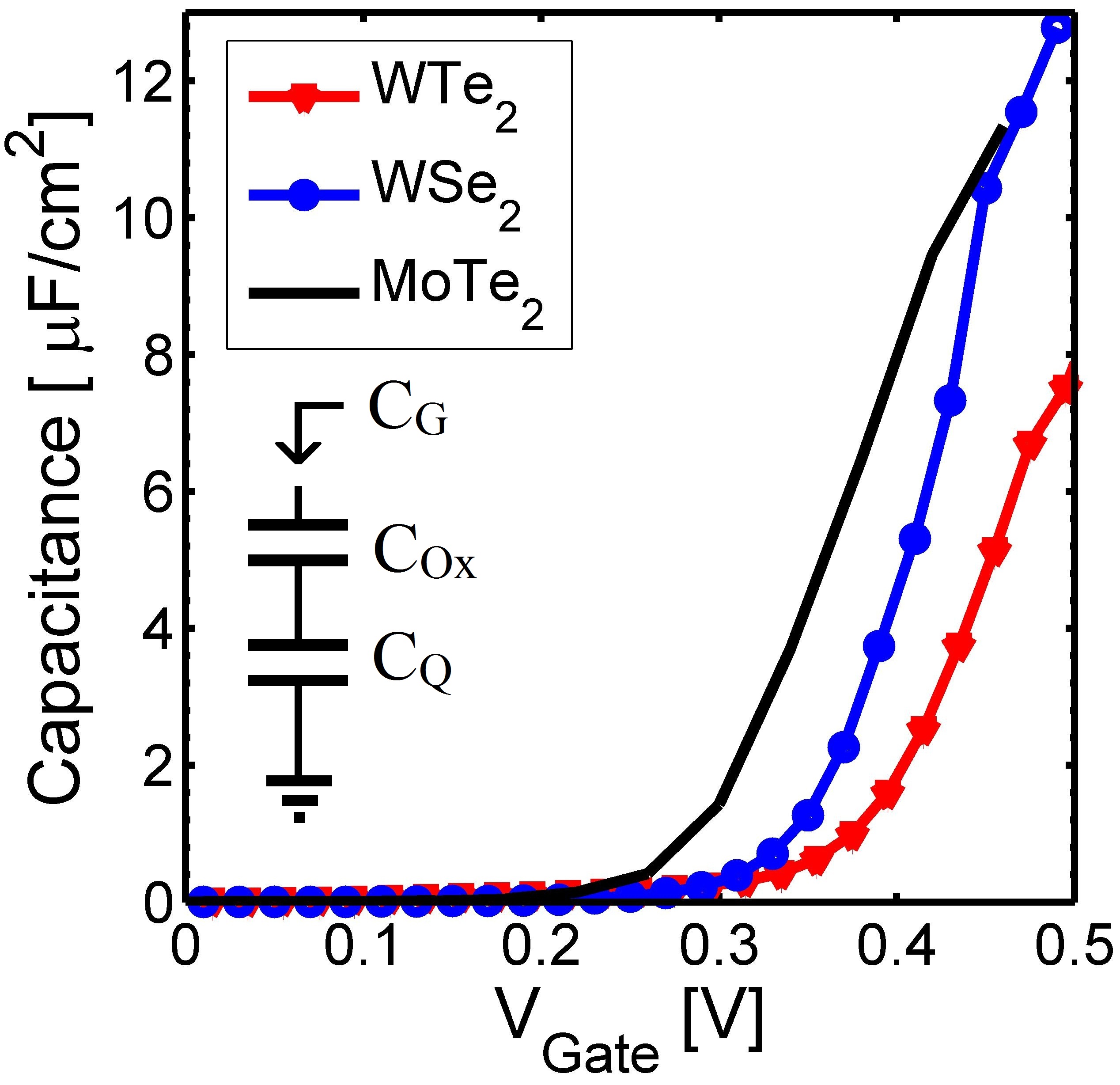}
                \label{fig:IdVg1}
        \end{subfigure}%
        \caption{C-V$_{Gate}$ of WTe$_2$, WSe$_2$, and MoTe$_2$ TFETs. }\label{fig:CV}
\end{figure}

Fig. \ref{fig:CV} shows the total gate capacitance versus the gate voltage (C-V) for the TMD TFETs under investigation. There are two major factors which determine the C-V characteristics: the quantum capacitance, and the threshold voltage. The total gate capacitance (C$\rm _G$) can be modeled as a series of oxide capacitance (C$\rm _{Ox}$), and quantum capacitance (C$\rm _Q$). The quantum capacitance is proportional to the density of states and m$^*$ of the material \cite{CQ_dev, Cq_chen}. Materials with larger m$^*$ have larger C$\rm _Q$ which translates into larger C$\rm _G$. On the other hand, the threshold voltage determines the voltage where the charge starts to appear in the channel. Since the off-currents of TFETs have been fixed here, the threshold voltage of one material is different from the other. Clearly, WTe$\rm _2$ is showing the lowest C$\rm _G$ values, another benefit when it comes to benchmarking of various TMD devices as discussed below.

Drain induced barrier lowering (DIBL) is one of the most important short-channel effects for ultra-scaled transistors. DIBL causes a reduction of the threshold voltage by the drain voltage, and it is one of the commonly used criteria for indicating short-channel behavior \cite{t_TMDC}. In TFETs, the active region in which tunneling occurs is right at the source-to-channel interface, far from the drain contact. Moreover, in 2D TFETs, gate control is stronger due to a thin channel which suppresses short channel effects. Consequently, short channel effects are less significant in comparison with conventional FETs (n-i-n or p-i-p doped transistors). The numerical values of DIBL for TMD TFETs obtained in this work are listed in table III. Note that DIBL values of WSe$_2$ and WTe$_2$ TFETs are substantially smaller than the reported 80 mV/V DIBL of ultra-scaled MOSFETs \cite{ITRS2, ITRS1}. DIBL is calculated at the current level of 1$nA/\mu m$ from the change in the threshold voltage caused by varying V$_{DS}$ from 0.1V to 0.5V. 
\begin{equation}
\label{eq:DIBL}
DIBL=- \frac{V_{Th} {(V_{DS}=0.5)} -V_{Th} (V_{DS}=0.1)}{0.5-0.1} \\
\end{equation}

Among these TMD TFETs, WSe$_2$ and WTe$_2$ show a smaller DIBL compared to MoTe$_2$. This is because WSe$_2$ and WTe$_2$ have a smaller in-plane dielectric constant compared to MoTe$_2$ which reduces the electric field penetration from drain, and hence suppresses short channel effects. 

\begin{table}[h!]\center
\caption{\label{tab:dibl} {DIBL values for TMD TFETs}}
    \begin{tabular}{| l | l | l | l |}
    \hline
	Material & WTe$_2$ & WSe$_2$ & MoTe$_2$ \\ \hline	
	DIBL [mV/V] & 25 & 20 & 67\\
	\hline
    \end{tabular}
\end{table}

\subsection{Subthreshold swing and energy-delay product}

\begin{figure}[h!]
        \centering
        \begin{subfigure}[b]{0.24\textwidth}
                \includegraphics[width=\textwidth]{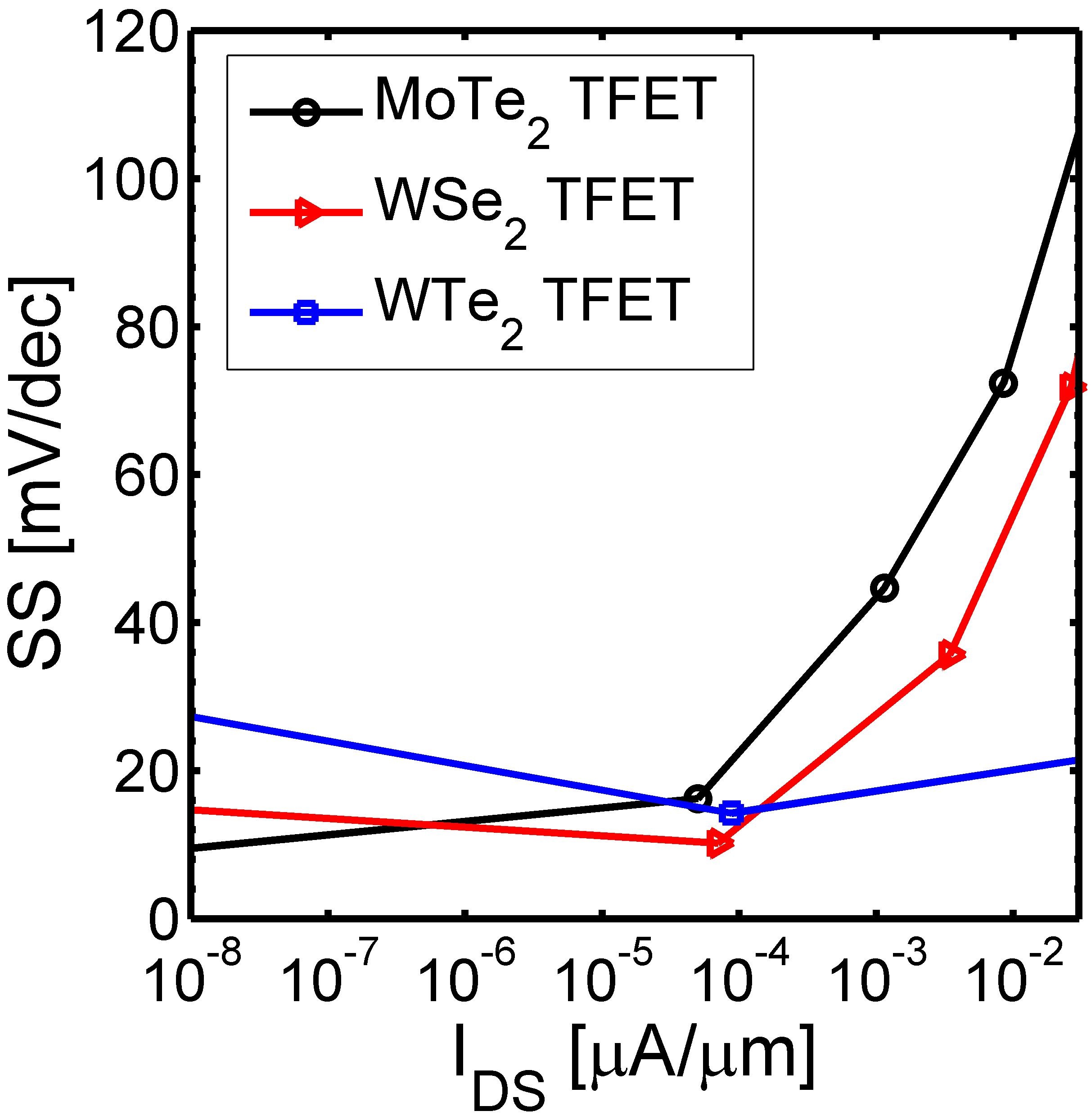}
                \caption{}
                \label{fig:SS}
        \end{subfigure}%
        ~ 
        \begin{subfigure}[b]{0.26\textwidth}
                \includegraphics[width=\textwidth]{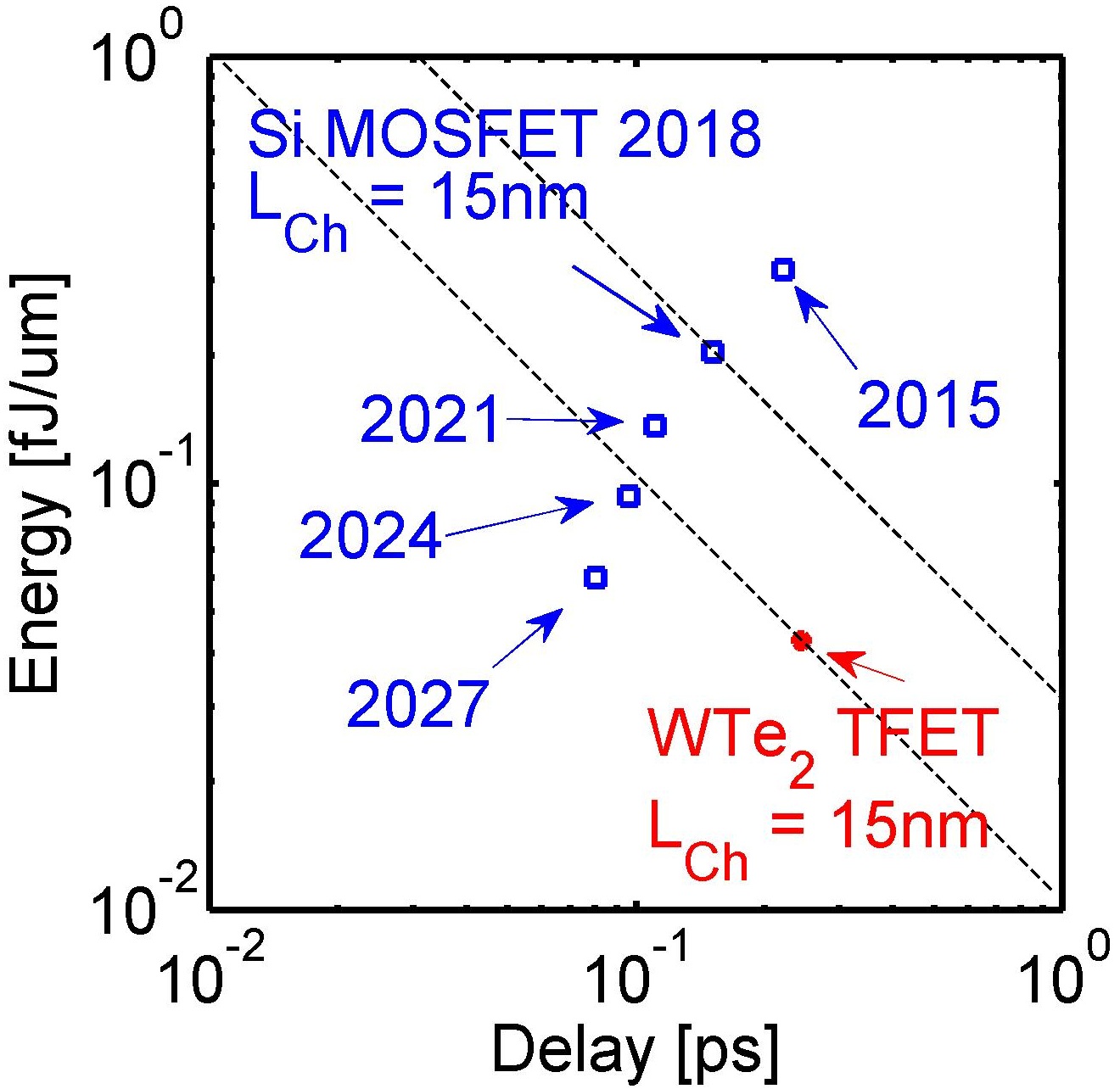}
                \caption{}
                \label{fig:E_tau}
        \end{subfigure}
        ~ 
        \caption{a) SS-I$\rm _{DS}$ for different TMD TFETs b) Energy-delay of WTe$_2$ TFET with a doping level of 2e20 cm-3 versus Si MOSFETs from ITRS \cite{ITRS2, ITRS1}. The dashed lines depict constant energy-delay products. }\label{fig:dop_effect}
\end{figure}

The main idea behind a TFET is to achieve a steep subthreshold swing below the Boltzmann limit of 60 mV/dec. It is important to compare the steep devices in a generic way. For example, the average subthreshold swing does not provide information about the current range in which the I-V is steep. Recently, a method for benchmarking steep devices has been proposed which gives insight into the local steepness of the I-V curve \cite{Seabaugh2}. In this method, the subthreshold swing is plotted against the drain-to-source current density of the device. Fig. \ref{fig:SS} shows how mono-layer TMD TFETs fit in this picture. Notice that the current on the horizontal axis is not the ON-current but the current where the SS has been calculated. The devices with the best performance will ideally be placed on the lower right corner of the figure, where the SS is small and the drain-to-source current is large. 

Fig. \ref{fig:SS} shows again that mono-layer WTe$_2$ is the most promising candidate for TFET applications among the TMD materials considered in this study. 
Fig. \ref{fig:E_tau} shows the energy ($C_{Gate} V_{DD}^2$) versus the intrinsic delay ($C_{Gate}V_{DD}/I_{ON}$) of a WTe$_2$ TFET versus ultra-scaled silicon MOSFETs. The WTe$_2$ TFET shows a smaller intrinsic energy-delay product value when compared to MOSFETs with the same channel length. In terms of the anticipated circuit performance of a 32-bit adder based on TMD TFETs, the performance specs discussed in this article for WTe$_2$ devices for $V_{DD}$=0.5V result in the similar energy-delay product as reported by Nikonov et al. \cite{Nikonov} for $V_{DD}$=0.25V. The actual energy and delay values were calculated to be 13fJ and 2450ps respectively using the same code as in \cite{Nikonov}.


\section{Conclusion}
In this work, the performance of various Transition Metal Dichalcogenide materials (MoS$_2$, WSe$_2$, MoTe$_2$, and WTe$_2$) TFETs has been investigated through self-consistent atomistic simulations. It has been shown that an atomically thin channel alone is not sufficient for high performance TFETs; especially to achieve high ON-currents, the choice of channel material and device design are critical as well. According to our analysis, WTe$_2$ is the most promising TMD in this study for TFET applications with high ON-currents of 350 $\mu A/\mu m$. TMD TFETs exhibit reduced short channel effects: DIBL and SS values are significantly lower than for Si MOSFETs (by a factor of about 1/3). Moreover, the energy-delay product of the optimized WTe$_2$ TFET is lower than that of an ultra-scaled Si MOSFET. Our simulations show that 2D materials with lower band gaps (0.5-0.7eV) and effective masses are more suitable for high performance TFETs.

\section*{Acknowledgment}
Authors would like to thank Joe Nahas and Robert Perricone for the 32 bit adder energy-delay calculations and Tarek Ameen for his help with the simulations. This work was supported in part by the Center for Low Energy Systems Technology (LEAST), one of six centers of STARnet, a Semiconductor Research Corporation program sponsored by MARCO and DARPA. The use of nanoHUB.org computational resources operated by the Network for Computational Nanotechnology funded by the US National Science Foundation under grant EEC-1227110, EEC-0228390, EEC-0634750, OCI-0438246, and OCI-0721680 is gratefully acknowledged.

\end{document}